\documentclass{kluwer}    

\newread\epsffilein    
\newif\ifepsffileok    
\newif\ifepsfbbfound   
\newif\ifepsfverbose   
\newdimen\epsfxsize    
\newdimen\epsfysize    
\newdimen\epsftsize    
\newdimen\epsfrsize    
\newdimen\epsftmp      
\newdimen\pspoints     
\def\epsfbox#1{\global\def\epsfllx{72}\global\def\epsflly{72}%
   \global\def\epsfurx{540}\global\def\epsfury{720}%
   \def\lbracket{[}\def\testit{#1}\ifx\testit\lbracket
   \let\next=\epsfgetlitbb\else\let\next=\epsfnormal\fi\next{#1}}%
\def\epsfgetlitbb#1#2 #3 #4 #5]#6{\epsfgrab #2 #3 #4 #5 .\\%
   \epsfsetgraph{#6}}%
\def\epsfnormal#1{\epsfgetbb{#1}\epsfsetgraph{#1}}%
\def\epsfgetbb#1{%
\openin\epsffilein=#1
\ifeof\epsffilein\errmessage{I couldn't open #1, will ignore it}\else
   {\epsffileoktrue \chardef\other=12
    \def\do##1{\catcode`##1=\other}\dospecials \catcode`\ =10
    \loop
       \read\epsffilein to \epsffileline
       \ifeof\epsffilein\epsffileokfalse\else
          \expandafter\epsfaux\epsffileline:. \\%
       \fi
   \ifepsffileok\repeat
   \ifepsfbbfound\else
    \ifepsfverbose\message{No bounding box comment in #1; using defaults}\fi\fi
   }\closein\epsffilein\fi}%
\def\epsfsetgraph#1{%
   \epsfrsize=\epsfury\pspoints
   \advance\epsfrsize by-\epsflly\pspoints
   \epsftsize=\epsfurx\pspoints
   \advance\epsftsize by-\epsfllx\pspoints
   \epsfxsize\epsfsize\epsftsize\epsfrsize
   \ifnum\epsfxsize=0 \ifnum\epsfysize=0
      \epsfxsize=\epsftsize \epsfysize=\epsfrsize
     \else\epsftmp=\epsftsize \divide\epsftmp\epsfrsize
       \epsfxsize=\epsfysize \multiply\epsfxsize\epsftmp
       \multiply\epsftmp\epsfrsize \advance\epsftsize-\epsftmp
       \epsftmp=\epsfysize
       \loop \advance\epsftsize\epsftsize \divide\epsftmp 2
       \ifnum\epsftmp>0
          \ifnum\epsftsize<\epsfrsize\else
             \advance\epsftsize-\epsfrsize \advance\epsfxsize\epsftmp \fi
       \repeat
     \fi
   \else\epsftmp=\epsfrsize \divide\epsftmp\epsftsize
     \epsfysize=\epsfxsize \multiply\epsfysize\epsftmp   
     \multiply\epsftmp\epsftsize \advance\epsfrsize-\epsftmp
     \epsftmp=\epsfxsize
     \loop \advance\epsfrsize\epsfrsize \divide\epsftmp 2
     \ifnum\epsftmp>0
        \ifnum\epsfrsize<\epsftsize\else
           \advance\epsfrsize-\epsftsize \advance\epsfysize\epsftmp \fi
     \repeat     
   \fi
   \ifepsfverbose\message{#1: width=\the\epsfxsize, height=\the\epsfysize}\fi
   \epsftmp=10\epsfxsize \divide\epsftmp\pspoints
   \vbox to\epsfysize{\vfil\hbox to\epsfxsize{%
      \includegraphics{#1}%
      \hfil}}%
\epsfxsize=0pt\epsfysize=0pt}%

{\catcode`\%=12 \global\let\epsfpercent=
\long\def\epsfaux#1#2:#3\\{\ifx#1\epsfpercent
   \def\testit{#2}\ifx\testit\epsfbblit
      \epsfgrab #3 . . . \\%
      \epsffileokfalse
      \global\epsfbbfoundtrue
   \fi\else\ifx#1\par\else\epsffileokfalse\fi\fi}%
\def\epsfgrab #1 #2 #3 #4 #5\\{%
   \global\def\epsfllx{#1}\ifx\epsfllx\empty
      \epsfgrab #2 #3 #4 #5 .\\\else
   \global\def\epsflly{#2}%
   \global\def\epsfurx{#3}\global\def\epsfury{#4}\fi}%
\def\epsfsize#1#2{\epsfxsize}
\let\epsffile=\epsfbox

\newdisplay{guess}{Conjecture}

\newcommand     \bfv  {{\bf v}}
\newcommand     \bbB  {\overline{\bf B}}

\newcommand     \ts  {\times}
\newcommand     \bfb  {\bf b}
\newcommand     \lb{\langle}
\newcommand     \rb{\rangle}
\newcommand     \curl{\nabla {\ts}}
\newcommand\bbJ{\overline {\bf J}}
\newcommand\bB{\overline { B}}

\newcommand\bfA{{\bf A}}

\newcommand\bfB{{\bf B}}
\newcommand\bfj{{\bf j}}
\newcommand\bfu{{\bf u}}
\newcommand\sm{{M_\odot}}

\newcommand\meanBB{\overline{\bf B}}
\newcommand\AAA{{\bf A}}
\newcommand\BB{{\bf B}}

\begin{document}                                                                                   
\begin{article}
\begin{opening}         
\title{Coronae \& Outflows from Helical Dynamos, Compatibility with the MRI, 
and Application to Protostellar Disks}
\author{Eric G. \surname{Blackman}}  
\institute{Dept. of Physics \& Astronomy, Univ. of Rochester,
Rochester NY, 14627, USA}

\author{Jonathan C. \surname{Tan}}  
\institute{Princeton Univ. Observatory, Peyton Hall, Princeton, NJ
08544, USA}
\runningauthor{Blackman \& Tan}
\runningtitle{Nonlinear Helical Dynamos as a Source of Outflows}


\begin{abstract}
Magnetically mediated disk outflows are a leading paradigm
to explain winds and jets in a variety of astrophysical sources, but 
where do the fields 
come from? Since accretion of mean magnetic flux may be disfavored in a thin
turbulent disk, and 
only fields generated with sufficiently large scale can escape 
before being shredded by turbulence, 
in situ field production is desirable. Nonlinear 
helical inverse dynamo theory can provide the desired fields for
coronae and outflows. We discuss the implications 
for contemporary protostellar disks, where the MRI (magneto-rotational
instability) can drive turbulence 
in the inner regions, and primordial protostellar disks,  where 
gravitational instability drives the turbulence.
We emphasize that helical dynamos are compatible with the magneto-rotational
instability, and clarify the relationship between the two.
\end{abstract}
\keywords{outflows, dynamos, accretion disk, magnetic fields, primordial
protstars}
\end{opening}           
\section{Introduction:  the dynamo-corona-outflow paradigm}

Bipolar jet-like outflows are commonly observed in protostellar systems  
(Richer et al. 2000),
as is enhanced X-ray activity
\cite{feigelson}. Though the correlation between outflows
and X-ray activity is not entirely systematic, 
magnetic fields probably play an important role for both phenomena.
Large scale outflows are likely magneto-centrifugally ``fling''
driven (Blandford \& Payne 1982) or magnetically ``spring''
driven (e.g. Lynden-Bell 1996) from large scale open field lines, 
while coronal dissipation is likely the result
of reconnection from closed loops, perhaps as they open (relax)
to form larger scale structures.  Both jet and coronal 
fields are plausibly the result of large scale fields produced
inside protostellar disks because 
(1) magnetic flux is difficult to accrete in
a thin turbulent disk (2)  only large scale fields are able to escape from the
disk before being shredded by turbulence therein.
A working paradigm is this: helical dynamo $\rightarrow$ 
large scale fields $\rightarrow$ coronal fields  
$\rightarrow$ field lines that relax and open up  
$\rightarrow$ magnetocentrifugally launched 
outflow with a significant supersonic vertical velocity
component $\rightarrow$ asymptotic outflow opening angle whose
tangent is the ratio of the expansion speed to the vertical
wind speed. The angle may be further reduced  
by magnetic collimation.

We focus here on understanding the principles that govern the strength
of large scale disk fields produced by dynamos, and on the mechanical
luminosities of the resulting winds.  We do not focus on the possible
role of stellar dynamos or the accretion ejection instability
(Tagger \& Pellat 1999; Varni\`ere \& Tagger 2002).


\section{Need for in situ generation of large scale fields}

\subsection{Small scale fields shred before they escape} 

For magnetic fields to launch jets or power coronae, the field buoyancy time
$t_b$, must be $<$ its diffusion time in the disk.  
Since $t_b \gsim h/{\overline U}_b$, where  $h$ is the disk 1/2
thickness and ${\overline U}_b$ is the buoyancy
speed ($\le$ Alfv\'en speed ${\overline U}_A$) 
associated with a  structure whose smallest gradient scale is $L$, 
we require 
$h/{\overline U}_b < L^2/\beta$  for escape, 
where
$\beta \sim u_2 l$ is the turbulent diffusion coefficient
for the magnetic field, 
and $u_2$ and $l$ are a measure of the dominant 
turbulent speed and scale, where  
(e.g. $l\sim $cube root of the turbulent cell volume $(l_x l_y l_z)^{1/3}$).
The escape condition can be rewritten 
$
L^2 >  l h (u_2/{\overline U}_b)
$ 
and since the ratio on the right (more on this later) is 
$\gsim u_2/{\overline U}_A \sim 1$,  
the condition $L^2 > l h$ is required for escape.


Now consider a Shakura-Sunyaev \cite{shakura73} disk viscosity 
$\nu_{ss}\sim \beta= u_{2} l \sim \alpha_{ss} 
c_s h \sim u_{2}^2 /\Omega \sim u_{A,2}^2/ \Omega,$
where $u_{A,2}$ is the Alfv\'en speed associated with the 
turbulent $B$-field and  $u_{2}\sim u_{A,2}$ follows
from simulations \cite{hgb95,bnst95}. 
Then $\alpha_{ss} \sim u_{A,2}^2/c_s^2$
or $u_{A,2} \sim \alpha_{ss}^{1/2} c_s$,
so that $l =\alpha_{ss}^{1/2}h$.
The escape condition $L^2>lh$ then  gives 
$L >  \alpha_{ss}^{1/4} h $.
Note that the minimum $L$ applies  to the 
{\it smallest} dimension of a given  structure 
since this determines the diffusion time.
This escape condition is not easily satisfied in the nonhelical
MRI dynamo (see minimum $k_y,k_z$ in fig. 3a (from Hawley et al. 1995)).
More on this point later. 


\subsection{Mean flux is not frozen: diffusion may beat accretion}

Can the large scale fields required by jets simply be accreted? 
Though 3-D MHD simulations are needed, 
it may be  difficult for thin disks.
The reason is that turbulent magnetic diffusivity $\beta\sim \nu_{ss}$
may keep the  large scale  field from being frozen on
an advection time scale. (In general,
the extent of flux freezing even for the total field 
depends on the turbulent spectrum. Turbulence
produces many small scale strucures which thus enhances the 
importance of the diffusion term 
in the induction equation.) 
To see the potential problem, we compare 
the ratio of the rate of flux advection to that of diffusion.  
Call $d$ the dominant vertical variation scale 
of the field strength.
The dominant variation scale of the radial velocity is 
$r$. The above mentioned ratio is then 
$\nabla\times({\overline {\bf v}}\times \bbB)/(\beta \nabla^2 \bbB)
\sim (d^2 / r^2)(\alpha_{ss} h c_s / \beta)$. 
The diffusion term thus dominates when the plausble 
(though not proven) relations $d < r$ and $\alpha_{ss} h c_s \sim \beta$ apply.
That diffusion can be faster than advection 
is supported by numerical 2-D calculation in \cite{lubow94}, but 
future 3-D numerical testing is  required.
This motivates an in situ helical dynamo,
whose growth rate can exceed the diffusion rate.

\section{MHD In Situ Amplification:  Direct vs. Inverse Dynamos}

We classify dynamos into two types: {\bf direct} and {\bf inverse}.
Direct dynamo action sustains magnetic fields on scales at or below 
the dominant turbulent scale.
This does not require helical turbulence although helicity 
does influence the magnetic spectrum. 
In contrast, inverse dynamo action describes  amplification 
on scales larger than that of the dominant scale of the turbulence.
The label ``inverse'' is used to suggest an inverse
cascade. The turbulence must be  helical 
to generate and sustain large scale flux over times longer than an
eddy turnover time which can in turn escape to coronae and drive outflows.
The distinction is illustrated in Fig 1. (Maron \& Blackman 2002)
for forced turbulence. The magnetic and kinetic energy spectra are plotted
for different values of the fractional helicity $f_h\equiv 
|\bfu_2\cdot\curl\bfu_2|/
|u_2^2 k_2|$.  The nonhelical forcing at $k_2=4.5$ produces no magnetic energy
at wavenumbers $k<k_2$, whereas the $f_h=1$ case produces a 
large peak at $k=1$.

\begin{figure}
\vspace{-.0cm} \hbox to \hsize{ \hfill \epsfxsize5cm
\epsffile{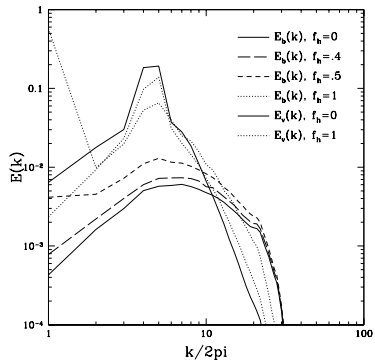} \hfill } 
\caption[]{Saturated dynamo energy spectra
(Maron \& Blackman 2002).}.
\end{figure} 

\subsection{Direct dynamo}

For nonhelical direct dynamo action, the field grows by random walk, 
field line stretching \cite{k68,parker79}.  
The turbulence can be 
externally driven by isotropic forcing 
(e.g. Maron \& Cowley 2002;
Haugen et al. 2003)
, or self-generated  
by an angular velocity gradient (e.g. Balbus \& Hawley 1998). In either case, 
turbulent stretching compensates for exponential
decay from turbulent diffusion. The latter operates
because the random motions of the gas also mix the field lines,
inducing a cascade to small scales where dissipation occurs. 
In 3-D, a steady state balance can be achieved.
In the saturated state, 
the magnetic energy $\sim$ turbulent kinetic energy 
integrated below the dominant turbulent scale.  
How far the magnetic energy peak is from the forcing scale 
and the overall nonhelical dynamo spectrum is an active area of research 
(e.g. Maron \& Cowley 2002; Schekochihin et al. 2002;
Haugen et al. 2003), which we do not discuss  further here.


\bigskip


\subsection{Helical inverse dynamo}

The helical inverse dynamo amplifies field on scales larger than that of
the turbulence.  This is  most desirable for coronae and outflows.
Fig 2a. shows a traditional $\alpha\Omega$  kinematic 
dynamo diagram  \cite{parker79}.
Consider an initially weak toroidal (=encircling the rotation axis)
loop of the magnetic field embedded in the rotator.
With a vertically decreasing density gradient,  
a rising turbulent eddy threaded by a magnetic field   
will twist oppositely to the underlying global rotation 
to conserve angular momentum.
Statistically, northern (southern) eddies  
twist field clockwise (counterclockwise). This is the ``$\alpha_d$'' 
effect and the result is a large scale poloidal field loop.
Differential rotation shears this 
loop (the ``$\Omega$''-effect). The bottom part 
reinforces the initial toroidal field and the top part diffuses
away. The result is exponential growth.
The $\alpha_d$ effect can be supplied by any source of 3-D turbulence
in a stratified rotator, including gravitational instability or the  MRI. 

This dynamo is revealed by averaging the magnetic induction 
equation over a local volume and breaking all quantities
(velocity $\bf U$, magnetic field $\bf B$ in Alfv\'en velocity
units, and  normalized current density ${\bf J}\equiv {\curl {\bf B}}$)
into their mean (indicated by an overbar) and 
fluctuating (lower case) components. The result is 
\begin{equation}
\partial_t\meanBB= \curl(\alpha_d\meanBB+{\overline {\bf U}}\ts \meanBB) 
+(\beta+\lambda) \nabla^2\meanBB,
\label{dynamo}
\end{equation}
%
where $\lambda$ is the magnetic diffusivity.
The ${\overline {\bf U}}$ 
term incorporates the $\Omega$-effect,
the $\beta$ term incorporates the turbulent diffusion 
(we assume constant $\beta$) 
and the first term on the right incorporates the $\alpha_d$-effect.
Accretion disk $\alpha-\Omega$ dynamo models have been 
proposed (Pudritz 1981; Reyes-Ruiz \& Stepinksi 1997;1999;
R\"udiger et al. 1995; Brandenburg \& Donner 1997; Campbell 1999; 
R\"udiger \& Pipin 2000) but without 
a fully dynamic theory of the field saturation value.
We now discuss this for forced turbulence, 
and expand further for disks in section 4.2.

\begin{figure}
\vspace{2cm}\hbox to \hsize{\hfill\epsfxsize12cm
\epsffile{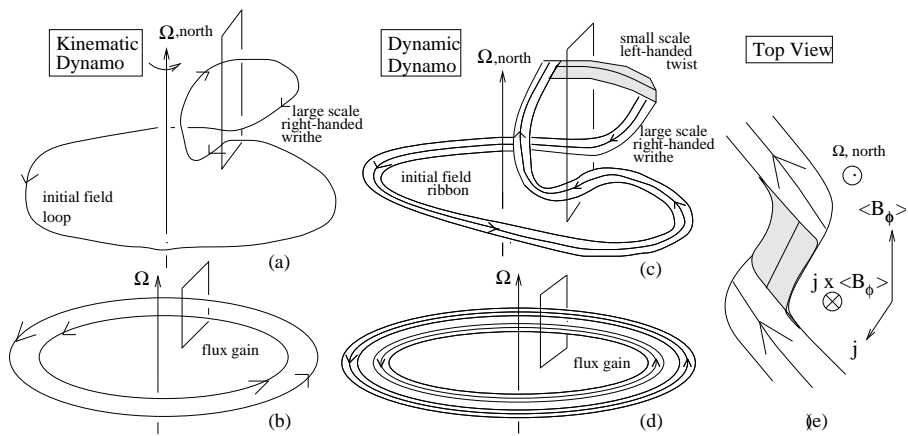} \hfill } 
\caption[]{Kinematic helical dynamo diagram (left)
ignores magnetic helicity conservation. In dynamic theory
field should be replaced by ribbon, showing the conservation  
(Blackman \& Brandenburg 2003).}
\end{figure}

For forced helical turbulence, 
$\alpha_d$ is  \cite{bf02} 
$\alpha_d=\alpha_{d0}+ \alpha_m$
where $\alpha_{d0}= -(\tau/3)
{\overline {\bfu\cdot\curl\bfu}}$, 
and $\alpha_m=(\tau/3){\overline {\bfb\cdot{\curl \bfb}}}$,
where $\tau$ is a turbulent damping time.
The $\alpha_m$ correction term arises from  
including the magnetic forces that backreact
on the velocity \cite{pfl,bf02}. 
From the form of $\alpha_d$, it is evident that 
growth of $\alpha_m$ can offset $\alpha_{d0}$ and quench the dynamo
\cite{fb02,bb02,bf02}.  
Magnetic helicity conservation determines
the dynamical suppression.
The magnetic helicity, a volume integral
$H\equiv \int\AAA\cdot\BB d V\equiv\lb\AAA\cdot\BB\rb V$ 
satisfies \cite{woltjer,berger84}
\begin{equation}
\partial_t H=
-2\lambda C-\mbox{surface terms},
\label{magcons}
\end{equation}
where $\nabla\times\AAA\equiv \BB$, 
and $C\equiv\lb{\bf J\cdot\bfB}\rb V$.
Without non-diffusive surface terms 
$H$ is typically well conserved in  astrophysics.
Since $H$ is  a measure of field line ``linkage'' and ``twist,'' 
its conservation implies that the $\alpha_d$-effect 
cannot produce a net magnetic twist, just positive and negative magnetic 
twists on different scales. 
The conservation equations like (\ref{magcons})
 for $H_1$ and $H_2$ 
(where $1(2)$ indicates large (small) scales)
are 
\begin{equation}
\partial_t H_1 = 2S
-2 \lambda k^2_1 H_1
-\mbox{surface terms} 
\label{h1}
\end{equation}
\begin{equation}
\partial_t H_2=-2S
-2\lambda {k^2_2 H_2}
-\mbox{surface terms},
\label{h2}
\end{equation}
where $S=(\alpha_d {k_1/\epsilon}-\beta {k}^2_1)H_1$ and 
$\alpha_d= \alpha_{0}+ \alpha_m$, 
where $\epsilon \le 1$ is the fraction of the large scale
field energy which is force-free.  
The solution without surface terms and with 
$\epsilon=1$ 
provides an estimate for the fully helical field large scale field 
component,  which can be further augmented by differential rotation. 
Because $\alpha_m\propto k_2^2 H_2$, the solution \cite{fb02,bb02,bf02} 
shows that for initially  small $H_2$ but large $\alpha_{0}$,
$H_1$ grows. Growth of $H_1$ implies the oppositely signed growth of $H_2$.
This $H_2$ backreacts on $\alpha_{0}$, 
ultimately quenching  $\alpha_d$ and the dynamo.

Fig. 2a does not capture magnetic helicity conservation
because a net helicity is generated. The resolution of this problem 
is  fig. 2b \cite{bb03}, 
where the field is treated as a ribbon. The large scale
writhe  corresponds to to the large scale magnetic helicity 
and to mean field dynamo action, while the small scale twist of opposite sense
along the ribbon is the result of magnetic helicity
conservation. The latter twist produces
a force which resists the bending of the field, and thus
encapsulates the backreaction and growth of $\alpha_m$.
Before discussing the saturation,  
more on the relation between the MRI and the direct vs. inverse 
helical dynamos is addressed.

\section{MRI is compatible with Direct and Inverse Dynamos}

\subsection{MRI in nonstratified disk: direct  dynamo only}

In a sufficiently ionized disk with 
a weak magnetic field (note: field must be strong enough for growth
to beat diffusion at critical wavenumber \cite{balbuspapaloizou})
the MRI results from an outward decreasing angular velocity gradient.
Without vertical stratification, and thus without 
any pseudoscalar helicity, the MRI does not produce magnetic fields on scales 
larger than that of the developed turbulence.
The marginally unstable (and dominant) growth
mode of the MRI satisfies $k u_A \sim \Omega$ \cite{balbus98}
so that as the magnetic field is amplified from the
instability, larger and larger scales
become unstable to growth. The largest scale
that can grow is that associated with the scale height $h$.
Without stratification or helicity, the largest scale of the turbulent magnetic
field matches the largest turbulent velocity scale 
(Fig. 3a \cite{hgb95}; $k_y$ is the toroidal wavenumber.)

The flux associated with the field
on the largest scale in the nonstratified MRI would 
change every turnover time as the field is shredded.   
The field is large scale only in the sense that 
the box height limits the scale of the turbulence NOT because
there is any sustained ordered flux on scales larger than the turbulence.  
Furthermore, for a periodic box, 
the net flux cannot grow, as it is conserved. 
Note also from Fig. 3a, that the largest field structure in the vertical $(k_z)$and radial $(k_x)$ 
directions are smaller than in the toroidal direction $(k_y)$.
This is important because 
the constraints of section 2.1 apply to the smallest dimension of
the largest scale structures. 
In sum, a nonstratified nonhelical 
MRI does {\it not} produce the fields required by jets/coronae.

Nonstratifed, nonhelical MRI driven turbulence 
and nonhelically forced turbulence behave  
qualitatively similarly with respect to the absence of 
fields on scales larger than the dominant scale of the turbulent motions.
Note however, that the nonhelical MRI,
$unlike$ the nonhelical forced case,
more nearly matches the turbulent kinetic energy spectrum on 
larger scales (compare the first two  panels in fig. 3 to the curves in fig. 1 
with $f_h=0$). This is likely due 
to the fact that the global shear growth time  is the same on all scales
in a shearing box, and thus more able to beat diffusion on the 
larger scales where it can provide additional magnetic energy 
compared to the case without global shear.

\subsection{MRI in stratified disk: both direct \& inverse dynamos}

Unlike the nonstratified case, 
when the  MRI operates in a vertically stratified 
medium it can have  a  built in inverse helical dynamo:
the stratification means that turbulence produced by 
the MRI  can be helical, in the manner discussed in section 3.2
and also from magnetic buoyancy (Brandenburg 1998; R\"udiger \& Pipin 2000). 
One can then use a nonlinear mean field theory to  model the growth
and saturation of the sustained large scale flux.
This is independent of the fact that the MRI will
always also  produce direct dynamo action that amplifies
fields as described. That is,
when the MRI operates in a stratified medium, 
one gets BOTH direct and inverse helical dynamo action.  

Mean field dynamo growth is  seen in magnetorotational
instability simulations (Brandenburg et al. 1995, 
Brandenburg \& Donner  1997).
Fig. 3b shows a dynamo cycle in which mean flux is maintained over
30 orbits. In that simulation, large and small scale  magnetic helicities are
seen to have opposite sign (as expected from discussion of section 3) 
but $\lb\bfu\cdot\curl \bfu\rb$ has the opposite 
sign to $\lb\bfj\cdot\bfb \rb$ which is not expected if 
$\alpha_{d}=\alpha_{d0}+\alpha_{m}$.
There is an additional 
buoyancy term 
$\alpha_b 
\sim - {\tau\over 3}|g_z\lb {\delta \rho\over \rho} b_x\rb|$ ($g_z$ is 
vertical gravity constant) which can exceed 
$\alpha_{d0}$  and carry 
the opposite sign (Brandenburg 1998; R\"udiger \& Pipin 2000).
Then, for example, in the northern hemisphere
$\alpha_d\sim -\tau |g_z\lb (\delta \rho/\rho) b_x\rb| +|\tau \lb \bfv\cdot \curl\bfv \rb|/3 +\tau \lb {\bfj\cdot \bfb }\rb /3$.
The 1st and 2nd terms on the right have opposite signs and 
the sign of the large scale magnetic helicity is determined by whichever
of those two initially dominates (since $\alpha_d$ is the growth coefficient).
This then predicts the sign of the last term on the right, which grows 
with opposite sign to the large scale magnetic helicity,
eventually quenching the larger of the 1st two terms on the right.


The fact that the stratfied MRI leads to both helical  dynamo
action and direct dynamo action, whereas the nonstratified
MRI leads only to direct dynamo action is the key to
clarifying previous discussions 
of the MRI and the tradititional mean field dynamo:
It is sometimes stated that even a nonstratified MRI generates 
``large scale'' fields without the need for helical turbulence.
But this statement relates directly to section 4.1.
The fields referred to as ``large scale'' in this nonhelical context
are just the result of direct (not helical) dynamo action:
1) they only reach the maximum scale of the turbulence, not
larger, and 2) they change the sign of their flux every eddy turnover time
(every rotation period). That such fields are generated without
helicity,  does not make the MRI inconsistent with helical dynamos: 
as discussed above, when stratified turbulence is presence
both direct dynamo and inverse helical dynamos operate together
and large scale fields on scales larger than that
of the dominant turbulence arise.

Like Brandenburg et al. (1995), Stone et al. (1996),
studied the stratified MRI. But the latter used 
vertically periodic boxes so no net flux could grow. 
Flux could have appeared in each ${1\over 2}$ 
of the Stone et al. (1996) box with opposite sign, 
but this was not a large region, given the resolution: 
Measuring finite pseudoscalars in a periodic stratified 
box requires averaging only over ${1\over 2}$  
of the vertical thickness, since the sign changes across the midplane.
For a periodic stratified box, the 
pseudoscalars must vanish at the midplane and 
at the top and bottom. 



\begin{figure}
\vspace{-.1cm} \hbox to \hsize{ \hfill \epsfxsize7cm
\epsffile{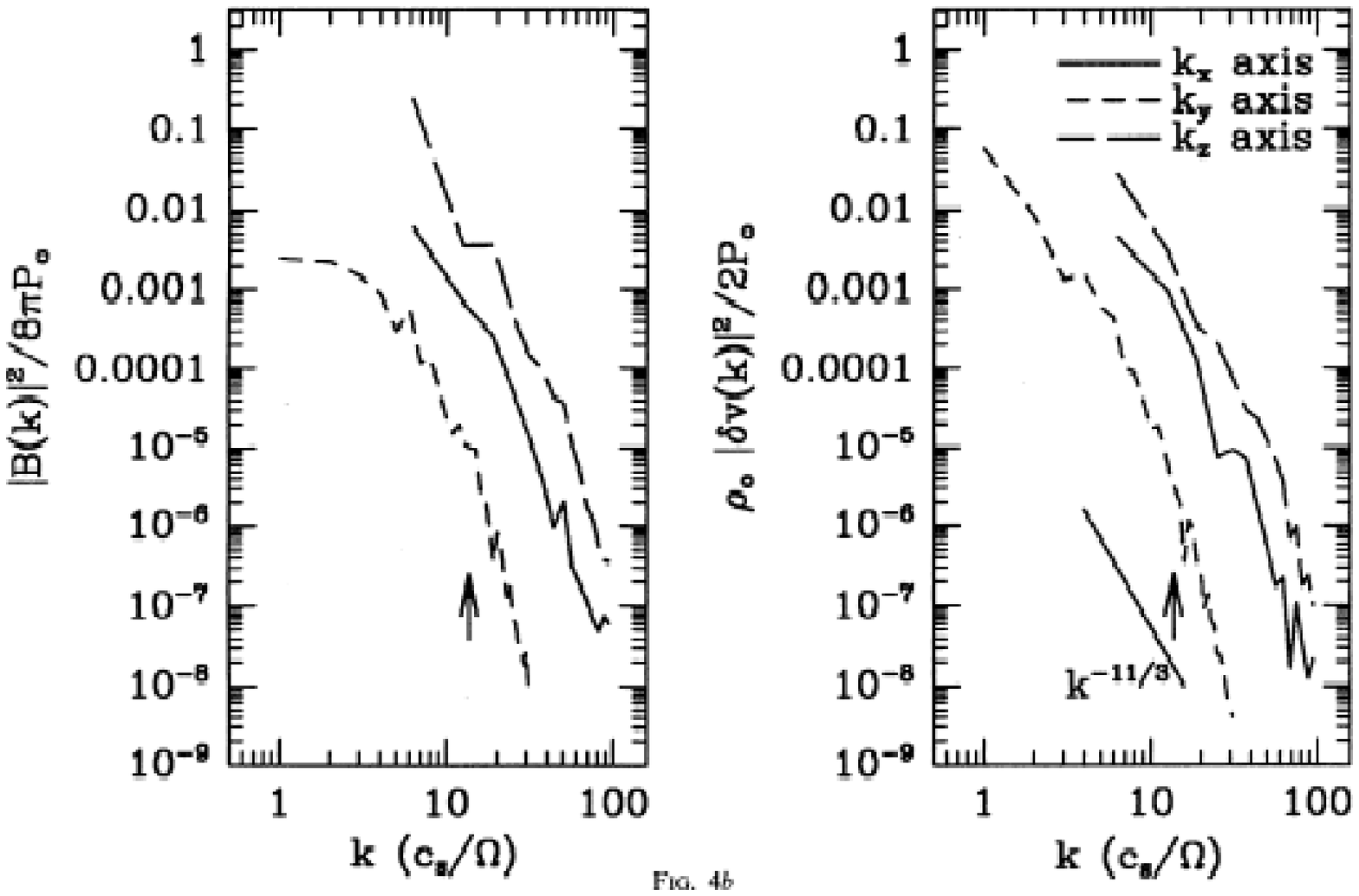} \epsfxsize4.5cm
\epsffile{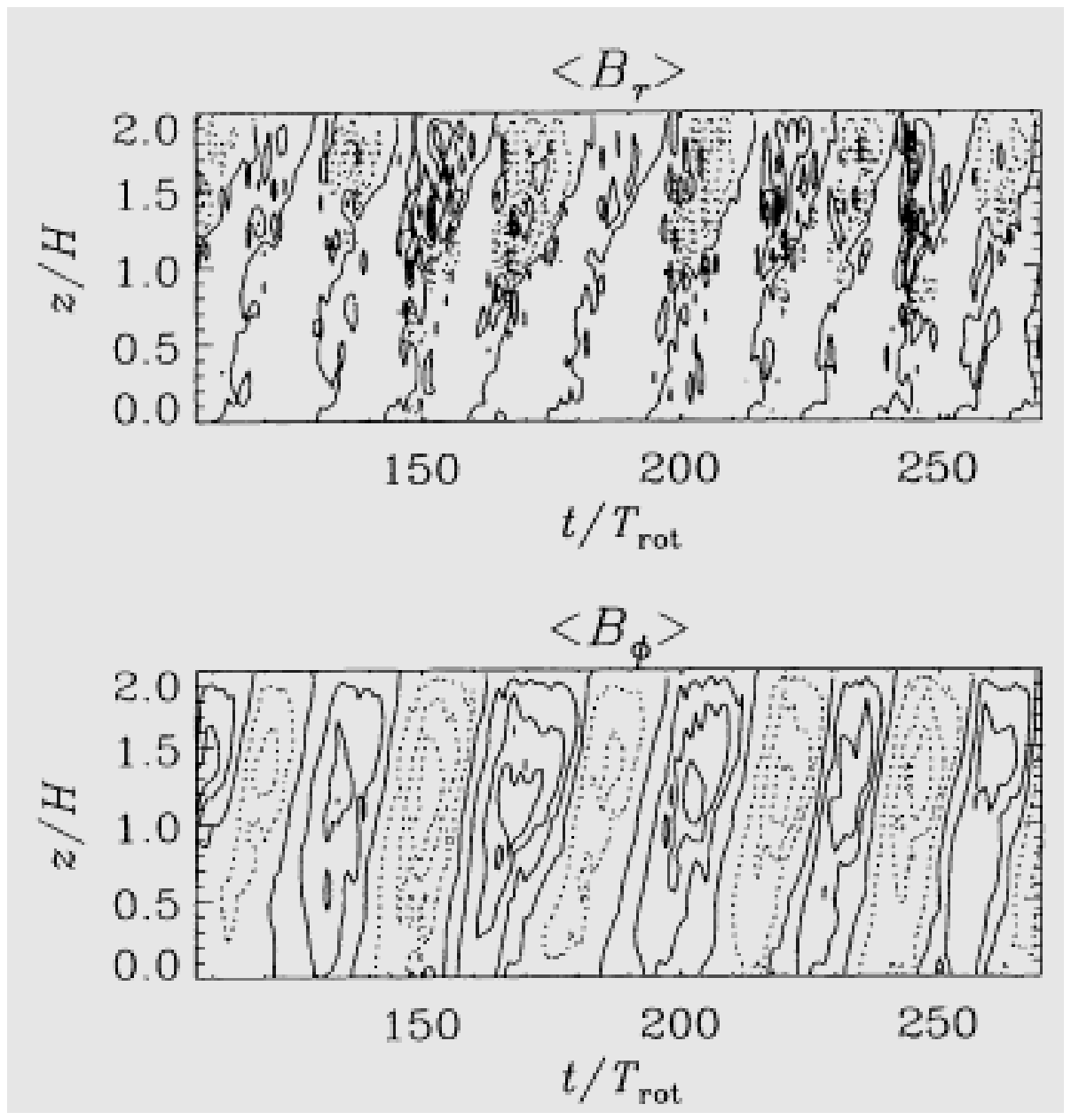} 
\hfill } \noindent
\caption[]{(a) Nonhelical  MRI spectra from Hawley et al. (1995)
and (b) MRI driven helical dynamo cycle in stratified 
non-periodic box (Brandenburg \& Donner  1997).}
\end{figure}

\section{Helical Dynamo Saturation in Disks \& Outflow Power}


The helical dynamo  grows helical large scale field 
first  ``kinematically'' (independent of
dissipation and with growth rate not too different from $\alpha_{d0}$)
until $|\alpha_m|\sim |\alpha_b-\alpha_{d0}|$, after which the field growth is 
resistively limited.  Setting  $|\alpha_m|\sim |\alpha_b-\alpha_{d0}|$,
and using magnetic helicity conservation, 
then provides a reasonable estimate of the helical field
by the end of the kinematic phase.  
Growth beyond that value
is often too slow to be of interest.  The large scale helical field
at the end of the  kinematic phase 
is (Blackman \& Field 2002; Tan \& Blackman 2003)
\begin{equation}
\bB_H \sim (4\pi \rho f_h)^{1/2} u_2(l/h)^{1/2} \simeq 
 (4\pi \rho)^{1/2} \alpha_{ss} c_s,
\label{sat}
\end{equation}
where $f_h$ is the fractional kinetic helicity,
$l$ is the dominant turbulent scale, 
$h$ is the disk
height (=the vertical scale of the growing mean field),
and the latter similarity follows for disks since 
$l/h\sim \alpha_{ss}^{1/2}$ and $u_2\sim \alpha_{ss}^{1/2}c_s$,
which in turn follow from the scalings of section 2.1.

Eqn. (\ref{sat}) is  an estimate of the 
helical field but 
the nonhelical toroidal field is further amplified 
by  shear. The toroidal field is linearly stretched above $\bB_r$ 
during a buoyancy 
time scale.
The buoyancy loss time  
is $\tau_{b}\sim  h (4\pi \rho)^{1/2}/\bB_\phi$, so  linear 
growth of  $\bB_\phi$ above (\ref{sat})
in a time $\tau_{b}$ gives 
$\bB_\phi=\bB_r \Omega \tau_{b}=\alpha_{ss}^{1/2}c_s (4\pi\rho)^{1/2}$
so 
$\bB_r/\bB_\phi= \alpha_{ss}^{1/2}$.

The above mean fields have their smallest variation 
scale $\sim h$  and thus satisfy the conditions of section 2.1
and can rise to the corona, open up, and launch outflows.
By using $\nabla\cdot {\overline {\bf B}}$ one can use 
the disk field to estimate the field at the surface  
and the Poynting flux there (Tan \& Blackman 2003). 
This provides an estimate for the magnetic luminosity 
available for magneto-centrifugal (Blandford \& Payne 1982)
 or magnetic ``spring'' (Lynden-Bell 1996)
driven outflows
\begin{equation}
\begin{array}{r}
L_{mag} \sim {c\over 4\pi}\int ({\overline {\bf E}}\ts {\overline {\bf B}})
\cdot d{\bf S}
\sim  \frac{\alpha_{ss}^{1/2}Gm_* \dot{m}_*}{r} \\
\simeq 220 \:L_\odot
\left(\frac{\alpha_{ss}}{10^{-2}}\right)^{1\over 2} \frac{m_*}{M_\odot}
\frac{{\dot m}_*}{10^{-2}M_\odot {/\rm yr}}\left({10^{13}{\rm cm}\over r}\right),
\label{lum2}
\end{array}
\end{equation}
using normalizations for primordial protostars (Tan \& McKee 2003). 



\section{Outflows in Contemporary \& Primordial Protostars}

\begin{figure}
\vspace{-.1cm} \hbox to \hsize{ \hfill \epsfxsize5cm
\epsffile{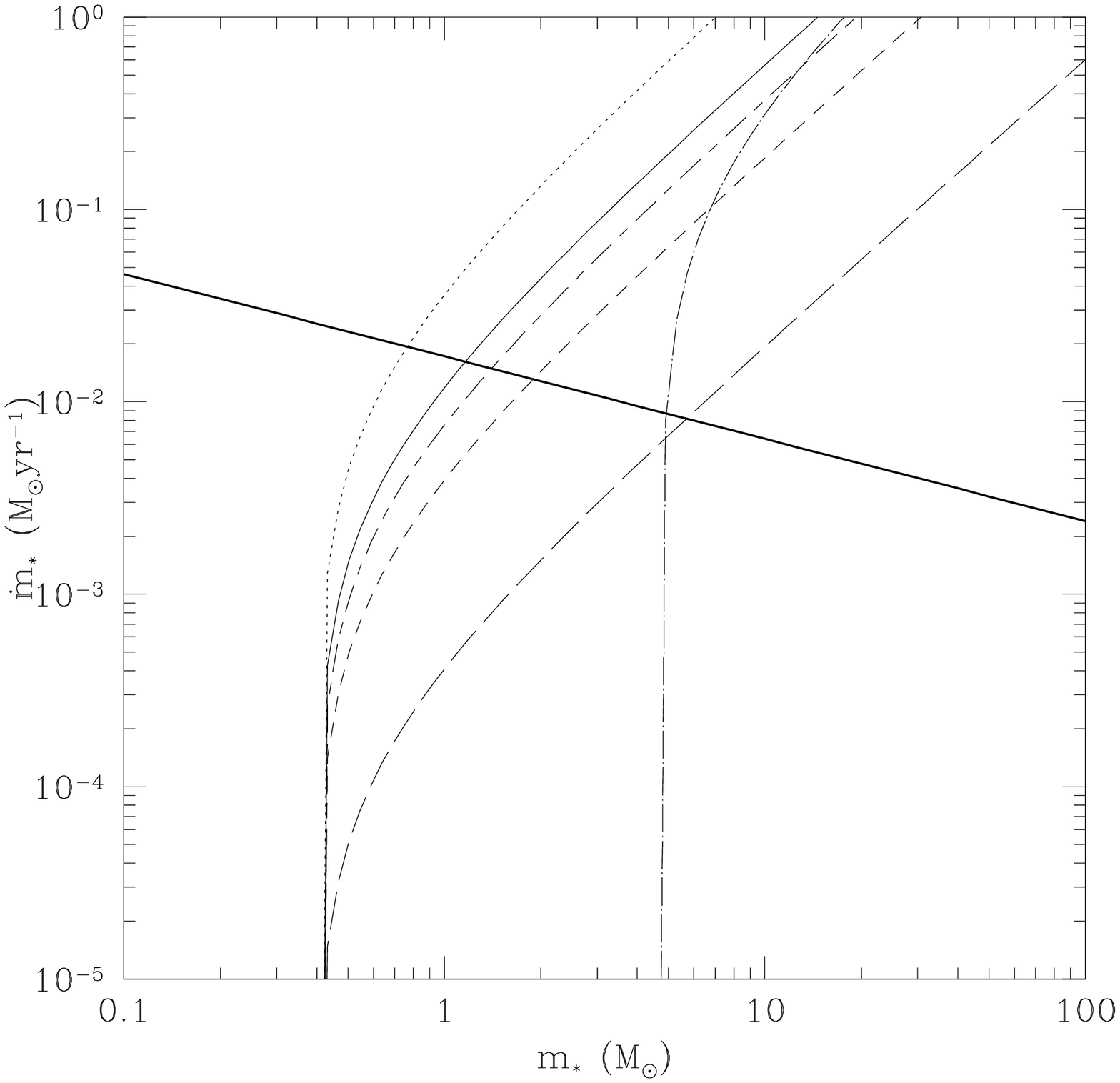} \epsfxsize5cm \epsffile{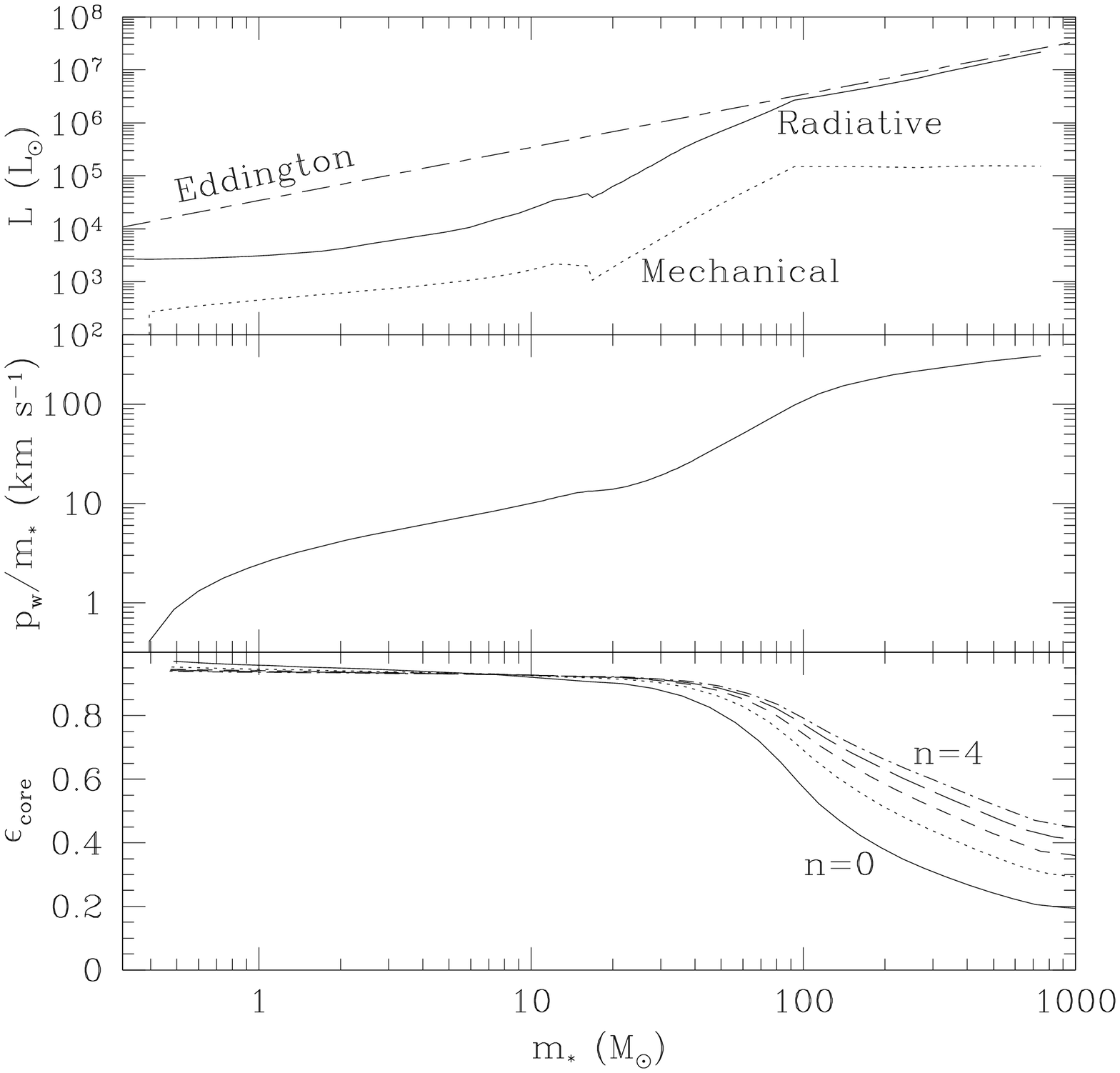} \hfill }  
\caption[]{(a) Dynamo saturation regime for primordial
protostellar disks (b) Evolution of radiative and  
wind mechanical luminosities from protostar and disk;
outflow momentum; effect of wind on star formation efficiency 
(Tan \& Blackman 2003).}
\end{figure}

The time scale for which the helical dynamo operates at a given radius
is not the viscous time, but the accretion disk's lifetime 
($\sim$ star formation time).  
This is because the mean field is NOT frozen into the
flow.  
The amplification is statistical, lasting as long as turbulence
is supplied.  For
contemporary star formation, we expect that growth to the kinematic
saturation value occurs in a small fraction of the star formation
time, since the initial fields after collapse from molecular clouds
are already quite strong and the star formation time scales are
relatively long. Contemporary protostellar disks should be able
to easily sustain outflows powered by fields from a helical dynamo,
with the helical turbulence in the inner regions of the disk supplied
by the MRI.

For primordial protostellar disks, 
although magnetic Reynolds numbers are high, the seed fields
are weaker and the MRI may be unfavorable 
for the maximally unstable mode. Instead, gravitational instability 
(Gammie 2001, Wada et al. 2002; though these are 2-D simulations) 
may supply the 3-D helical turbulence for much of
the dynamo evolution. In addition, we expect the star formation time
scales to be relatively short since the accretion rates are $\sim
10^{-2}M_\odot/{\rm yr}$ (Tan \& McKee 2003). 
Gravitational instability in 
the disk may maintain the Toomre parameter $Q\lsim
1$ (Gammie 2001; Tan \& Blackman 2003):  Where the disk is initially
stable (such that viscosity is small), mass  accumulates, 
triggering the instability. 
This can induce turbulent accretion rather than global angular momentum
transport (e.g. Gammie 2001). Typical accretion rates, while large by
contemporary standards, are small enough that fragmentation 
is not expected  (Tan \& Blackman 2003).

Tan \& Blackman (2003) used a primordial protostellar model (Tan
\& McKee 2003) to determine the $m_*$  vs. $\dot{m}_*$ regime for which a helical 
dynamo reaches the end of the kinematic regime (discussed above).  
In Fig. 4a., the rising
curves mark the condition where the helical field has
been able to saturate, as the disk is old enough to have allowed
sufficient dynamo amplification.  The disk dynamo is
assumed to start operating once the centrifugal radius of the
accretion flow is $\sim$ twice the stellar radius, 
which happens when $m_{\rm *,init}=0.4\sm$ in the fiducial case. 
The dashed, solid, and
dotted lines show the saturation condition for $\alpha_{\rm
  ss}=10^{-3},10^{-2},0.1$, all with radius $r=10^{13}\:{\rm cm}$ and seed
field of $10^{-16}\:{\rm G}$.
With $\alpha_{\rm ss}=0.01$, the long-dashed line shows the condition
at $r=10^{14}\:{\rm cm}$, the dashed-long-dashed line shows it for a
seed field of $10^{-26}\:{\rm G}$, and the dot-long-dashed line shows
it evaluated with $m_{\rm *,init}=4.7\sm$, corresponding to a more
slowly rotating initial gas core.

Protostellar growth starts from low masses with accretion rates that
decline with time (and so also with protostellar mass), as shown by
the straight thick solid line. In the fiducial case the field in the
inner disk saturates when $m_*\sim 1\sm$.  This does not depend 
sensitively on $\alpha_{\rm ss}$. In lower angular momentum cores the
disk emerges later and the dynamo then rapidly saturates.  
The main message of Fig 4a is that
dynamo saturation and thus dynamo-driven winds may occur at relatively
low protostellar masses, even in the primordial case.


 
Fig. 4b (Tan \& Blackman 2003) 
shows the evolution of protostellar
luminosities (radiative and mechanical), the cumulative specific
outflow momentum relative to stellar mass (assuming that the wind
speed is the escape speed from the protostellar surface and that
the flow kinetic energy dominates far from the
star), and the resulting star formation efficiency due to erosion of
the core by the outflow for the fiducial
case of Fig 4a, with $\alpha_{\rm ss}=0.01$.  The 
efficiency calculation is based on Matzner \& McKee (1999,2000).
The initial core's axisymmetric angular density profile is
specified by $dM/d\Omega =(1/4\pi) Q(\mu) M$, with $\mu = {\rm cos}
\theta$ and $Q(\mu)=(1-\mu^2)^n / \int_0^1 (1-\mu^2)^n \:d\mu$. The solid
line is $n=0$ (isotropic core), dotted is $n=1$, dashed is $n=2$,
long-dashed is $n=3$, dot-dashed is $n=4$.  The message from fig. 4b
is that dynamo driven winds can strongly affect the star
formation efficiency for masses $m_*\gtrsim 100\sm$, 
at which the protostar has contracted to the main sequence radius, 
and the contraction has enhanced  the 
accretion and wind luminosities.

\end{article}

\end{document}